\documentclass[sigconf]{acmart}
\AtBeginDocument{%
  }

\acmConference[MM '25] {Proceedings of the 33rd ACM International Conference on Multimedia}{October 27--31, 2025}{Dublin, Ireland.}
\acmBooktitle{Proceedings of the 33rd ACM International Conference on Multimedia (MM '25), October 27--31, 2025, Dublin, Ireland}
\copyrightyear{2025}
\acmYear{2025}
\setcopyright{acmlicensed}

\acmISBN{979-8-4007-2035-2/2025/10}
\acmDOI{10.1145/3746027.3754828}

\usepackage{tabularx}
\usepackage{multirow}
\usepackage{multicol}
\usepackage{tcolorbox}
\usepackage{subfig,graphicx}
\usepackage{fontawesome5}

\begin{document}

\title{From Individuals to Crowds: \\Dual-Level Public Response Prediction in Social Media}

\author{Jinghui Zhang}
\orcid{0009-0003-6403-1113}
\affiliation{%
  \institution{Mohamed bin Zayed University of Artificial Intelligence}
  \city{Abu Dhabi}
  \country{United Arab Emirates}}
\email{jinghui.zhang@mbzuai.ac.ae}

\author{Kaiyang Wan}
\orcid{0000-0001-8669-2551}
\affiliation{%
  \institution{Mohamed bin Zayed University of Artificial Intelligence}
  \city{Abu Dhabi}
  \country{United Arab Emirates}}
\email{kaiyang.wan@mbzuai.ac.ae}

\author{Longwei Xu}
\orcid{0009-0005-6131-0015}
\affiliation{%
 \institution{Shandong University}
 \city{Qingdao}
 \state{Shandong}
 \country{China}}
 \email{longwayxu@mail.sdu.edu.cn}

 \author{Ao Li}
 \orcid{0009-0008-4560-770X}
\affiliation{%
 \institution{Shandong University}
 \city{Qingdao}
 \state{Shandong}
 \country{China}}
 \email{liaolea@mail.sdu.edu.cn}

\author{Zongfang Liu}
\orcid{0009-0003-3153-3010}
\affiliation{%
  \institution{Mohamed bin Zayed University of Artificial Intelligence}
  \city{Abu Dhabi}
  \country{United Arab Emirates}}
\email{zongfang.liu@mbzuai.ac.ae}

\author{Xiuying Chen}
\orcid{0000-0002-6633-0796}
\authornote{Corresponding author.}
\affiliation{%
  \institution{Mohamed bin Zayed University of Artificial Intelligence}
  \city{Abu Dhabi}
  \country{United Arab Emirates}}
\email{xiuying.chen@mbzuai.ac.ae}

\renewcommand{\shortauthors}{Jinghui Zhang et al.}

\renewcommand{\arraystretch}{0.9}
% \addtolength{\textfloatsep}{-0.0cm}
\setlength\textfloatsep{0.5cm}
% \addtolength{\dbltextfloatsep}{-0.0cm}
\setlength\dbltextfloatsep{0.1cm}
% \addtolength{\abovecaptionskip}{-0.0cm}
\setlength\abovecaptionskip{0.3cm}

% \addtolength{\belowcaptionskip}{-0.0cm}
\setlength\belowcaptionskip{0.3cm}

\begin{abstract}
Public response prediction is critical for understanding how individuals or groups might react to specific events, policies, or social phenomena, making it highly valuable for crisis management, policy-making, and social media analysis. 
However, existing works face notable limitations.
First, they lack micro-level personalization, producing generic responses that ignore individual user preferences. 
Moreover, they overlook macro-level sentiment distribution and only deal with individual-level sentiment, constraining them from analyzing broader societal trends and group sentiment dynamics.
To address these challenges, we propose SocialAlign, a unified framework that predicts real-world responses at both micro and macro levels in social contexts. 
At the micro level, SocialAlign employs SocialLLM with an articulate Personalized Analyze-Compose LoRA (PAC-LoRA) structure, which deploys specialized expert modules for content analysis and response generation across diverse topics and user profiles, enabling the generation of personalized comments with corresponding sentiments.
At the macro level, it models group sentiment distributions and aligns predictions with real-world sentiment trends derived from social media data. 
To evaluate SocialAlign in real-world scenarios, we introduce SentiWeibo, a large-scale dataset curated from authentic social interactions on the Weibo platform. 
Experimental results on our SentiWeibo and related LaMP benchmark demonstrate that SocialAlign surpasses strong baselines, showing improved accuracy, interpretability, and generalization in public response prediction.
We hope our work inspires further research in public response prediction and computational social science: \url{https://github.com/Znull-1220/SocialAlign}.
% \faGithub ~\href{https://github.com/Znull-1220/SocialAlign}{here}.

\vspace{-0.1cm}
\end{abstract}


% \ccsdesc[300]{Computing methodologies~Natural language processing}
% \ccsdesc[100]{Applied computing~Sociology}


% Figure


\maketitle

\section{Introduction}

\begin{figure*}[tb]
    \centering
    \includegraphics[width=0.95\linewidth]{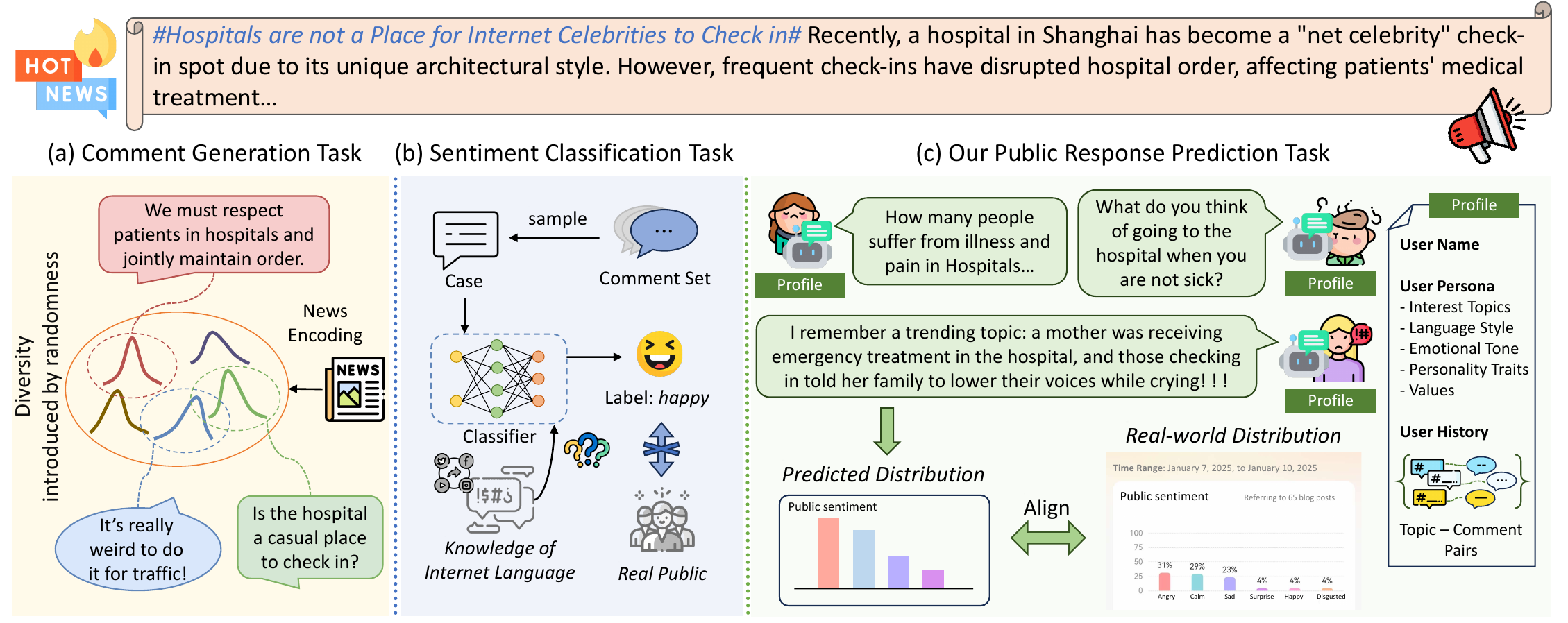}
    \caption{
(a) Existing models rely on randomness for diversity, ignoring real user personas.
(b) Traditional sentiment analysis treats comments independently, missing social interdependencies.
(c) SocialAlign integrates user profiles to generate personalized comments and analyze sentiment at both individual and societal levels, aligning with real-world trends.
    }
    \label{fig:intro}
\end{figure*}

Social media has become a key platform for sharing information and expressing emotions, especially during breaking news or major events~\citep{ausat2023theroleofsocialmedia}. The behavioral data gathered here offers unprecedented insights into public reactions,  group emotions, and broader sentiment dynamics~\citep{stefan2013DiffusioninSocialMedia}. Accurately predicting public response through micro-level comment analysis and macro-level sentiment distribution is crucial for understanding social behavior.
enabling crisis intervention, policy-making, and data-driven  decisions~\cite{gori1993social,jordan2018using}.

Despite progress in public response prediction, current methods struggle to capture complex social reactions. As shown in Figure~\ref{fig:intro}(a), conventional comment generation models~\citep{yang2019read,wang2021generating} rely on randomness for diversity but ignore micro-level personalization, leading to responses that miss users’ true sentiments. Meanwhile, sentiment analysis~\citep{fan2022sentencelevel} in Figure~\ref{fig:intro}(b), focuses on individual comments and lacks macro-level insight into societal trends. These gaps highlight the need for a unified framework combining personalized response generation with group-level sentiment modeling for a more accurate and interpretable understanding of public opinion.

To address the challenge of comprehensive public response prediction, we propose SocialAlign, a framework that integrates micro-level response generation with macro-level sentiment analysis, as illustrated in Figure~\ref{fig:intro}(c). 
At the micro level, SocialAlign retrieves relevant posts from user histories and constructs personalized personas for each individual based on their historical interactions.
These personas are defined along five dimensions: interests, language style, emotional tone, personality traits, and values. 
These personas inform SocialLLM, a specialized model designed to generate user-specific response comments. 
SocialLLM leverages a novel Personalized Analyze-Compose LoRA (PAC-LoRA) architecture, which incorporates expert modules tailored for content analysis and response generation across diverse topics and user profiles. 
This expert-driven approach not only improves the accuracy and personalization of generated responses but also ensures transparency by enabling traceable insights into the model's decision-making process.
At the macro level, SocialAlign aggregates individual responses to predict overall sentiment distributions, effectively capturing large-scale public opinion dynamics. 
This dual-level approach overcomes the limitations of individual-focused analysis methods by providing a comprehensive view of collective social responses.
% By bridging micro-level insights and macro-level sentiment trends, SocialAlign provides a powerful tool for understanding and anticipating collective behaviors.

To evaluate SocialAlign in real-world settings, we construct a dataset from organic social interactions on Weibo, comprising 476,605 user posts and sentiment distributions across 7,837 hashtagged events, 53 topics, and 6,401 users.  
Our evaluation spans both micro and macro levels. At the micro level, we assess generated comments via text quality and sentiment accuracy, introducing personalized evaluation metrics for both LLM and human assessments.  
At the macro level, we use Jensen-Shannon divergence to measure alignment between predicted and actual sentiment trends. 
Results on our dataset and related LaMP benchmark~\cite{chen2024large} show that SocialAlign and its PAC-LoRA component consistently outperforms both vanilla LLMs and fine-tuned baselines across all metrics.

Our contributions can be summarized as follows: 

\noindent $\bullet$ We establish a novel dual-level public response prediction paradigm that unifies micro-level response generation with macro-level sentiment distribution modeling, addressing a fundamental limitation in existing approaches that treat these aspects separately.

\noindent $\bullet$ We propose the SocialAlign framework, which effectively aligns language model outputs with real-world sentiment distribution at micro and macro levels, providing a practical and interpretable tool for social sentiment analysis. 

\noindent $\bullet$ We construct a large-scale dataset from real-world social interactions that addresses a critical gap in existing resources by integrating personalization and public sentiment alignment.

\section{Related Work}

\textbf{Public Response Prediction.} 
% comment generation
% Logic: importance -> response without personalization
% LLM prediction -> personalization -> flaws
Existing works on public response prediction include comment generation and sentiment analysis.
Early comment generation methods introduced randomness—e.g., via VAEs~\citep{chan2021enhancing,wang2021generating}—to enhance diversity.
\cite{zeng2019automatic} models user personalities through user embeddings and gated memory, while recent LLM-based methods like~\citep{nan2024let} simulate silent users to produce pseudo comments.
For sentiment analysis, traditional methods rely on lexicons~\citep{moreno-ortiz2017lingmotif} or classifiers like Naive Bayes~\citep{Kang2012NaïveBayes}, whereas recent models~\citep{hu2022unimse,AnandSidharth2023Multi-label} integrate multi-modal data for better predictions.
\cite{yu2024popalm} fine-tunes LLMs via reinforcement learning to generate trending responses.
Unlike prior work focused on either generation or sentiment analysis, our approach combines personalized generation with macro-level sentiment prediction, aligning LLM outputs with real-world public opinion for a fuller view of social response.

\textbf{Personalized LLMs.} 
Personalization is key for LLMs to adapt to user preferences and predict real-world trends. Existing approaches fall into two categories: prompt-based\citep{kang2023llmsunderstand,wang2024learningpersonalized} and finetune-based\citep{tan2024democratizing,yu2024neeko}.
Prompt-based methods guide generation by encoding user history as contextual examples~\citep{kang2023llmsunderstand,wang2024learningpersonalized}; for instance, \citep{mysore2024pearl} retrieves user-authored documents to augment prompts, while \citep{kang2023llmsunderstand} fine-tunes LLMs to predict user ratings.
Finetune-based methods directly encode user preferences into LLMs. \citep{tan2024democratizing} introduces PEFT modules to capture user-specific behavior, while others explore generalization and efficiency via MoE-style gating~\citep{yu2024neeko}, parameter merging~\citep{jang2023personalizedsoup}, and iterative learning~\citep{li2023teachllmspersonalize}.
Building on this, our work uses personalization to predict public responses, offering new tools for targeted communication, sentiment analysis, and policy-making.

% By integrating user-specific insights into both micro-level and macro-level predictions, we aim to enhance the effectiveness and relevance of public response modeling.
%\citet{yu2024neeko} propose to learn personalization in multi-character role-playing through gating mechanisms  这个MoE-like
%To tackle the scalability challenges with massive user data, works like PEARL~\cite{mysore2024pearl} employs retrieval augmentation to select suitable user historical items.
% Beyond direct retrieval, summarizing user history into high-level user profiles is also a promising approach, demonstrating strong enhancement in role-playing capabilities of LLMs. 
% In parallel, PEFT-based personalization fine-tunes a small number of parameters of LLM to learn individual preferences and behavior patterns. 

\textbf{Sentiment Analysis.} 
Sentiment analysis has evolved from the positive-negative binary classification~\cite{liu2021MLPs} to capturing contextual attitudes~\cite{chatterjee2019semeval}.
%~\cite{alhuzali2021spanemo} multilabel
Traditional approaches rely on sentiment lexicon identification~\cite{moreno-ortiz2017lingmotif} or machine learning algorithms such as Naive Bayes~\cite{Kang2012NaïveBayes}, while recent models~\cite{AnandSidharth2023Multi-label,li2025emoverse} have achieved state-of-the-art performance through fusing multi-modal information.
However, these methods typically neglect the social context where people express sentiments, a scenario that is essential to analyze real public emotions, especially on social media platforms~\cite{SANCHEZRADA2019socialcontext}.
Our work bridges this gap by extending sentiment analysis beyond the micro individual level to the macro group level, where we predict the collective sentiment distribution across a group of users.
% characterizing both individual nuances and social sentiment trends.

\textbf{Public Response Dataset.}
There are several publicly available benchmark datasets for public response prediction. 
Early datasets like Reddit Conversations and Youtube Comments focus on the diversity of response generation, but lack user-specific context.
Recent works have attempted to incorporate personal information to model individuals more accurately, but still face critical limitations. 
Although datasets like PersonaChat~\cite{zhang2018personalizing} have pioneered personality-aware response modeling, they rely heavily on synthetic personality descriptions that fail to capture real-world social dynamics. 
% This disconnect from authentic social contexts restricts their applicability for studying broader societal concerns and public sentiment analysis.
Even works like DAZE~\cite{sun2022JointlyModeling} using real-world data lack rich user features, as limited inputs like posts and avatars fail to support fine-grained sentiment prediction or personalized language modeling.
These observations motivate us to construct the SentiWeibo dataset, which captures public responses to social events while preserving user information.

\section{Problem Formulation}
In this section, we formalize the key concepts and notation used throughout our work. 
Formally, for each user $u$, there are $N_u$ historical posts, denoted as:
\begin{equation}
    \mathcal{H}_u = \{(x_u^1, y_u^1), (x_u^2, y_u^2), \cdots, (x_u^{N_u}, y_u^{N_u})\},
\end{equation}
where $x_u^i$ represents the topic (hashtag) of the $i$-th post, and $y_u^i$ denotes the corresponding user response content.

At the micro level, given a new topic $x^{\mathcal{N}}$, our objective is to generate a response $y_u^{\prime \mathcal{N}}$ that aligns with the ground truth response $y_u^{\mathcal{N}}$ in semantic content and language style. For this topic, we collect $M$ distinct user responses $\{y_1^{\prime \mathcal{N}}, y_2^{\prime \mathcal{N}}, \cdots, y_M^{\prime \mathcal{N}}\}$, each associated with a sentiment value $s_i$ where $i \in \{1,\dots,M\}$. 
At the macro level, these individual sentiment values collectively form a sentiment distribution $P^{\prime}$. Our objective is to minimize the JS divergence between the predicted distribution $P^{\prime}$ and the ground truth distribution $P$.

\section{Dataset Construction}
\begin{figure}[tbp]
    \centering
    \includegraphics[width=0.9\linewidth]{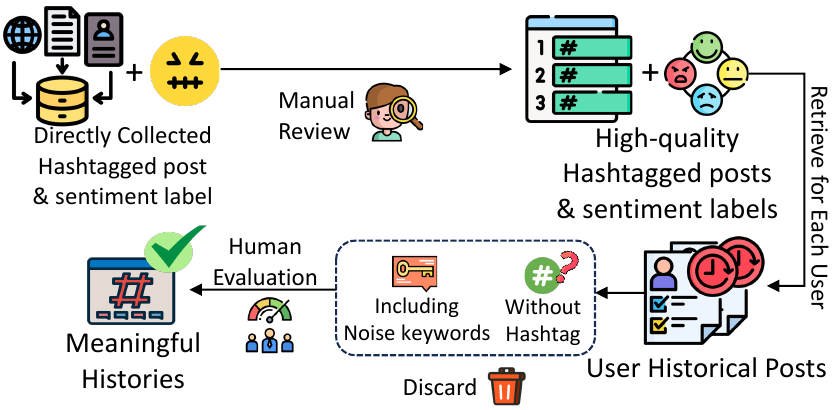}
    \caption{Data collection and pre-processing pipeline of our SentiWeibo dataset, which includes the comment content and corresponding sentiment labels.}
    \label{fig:dataset_pipeline}
\end{figure}

To overcome prior limitations, we build SentiWeibo from real Weibo interactions (Figure~\ref{fig:dataset_pipeline}).

\subsection{Data Collection}

Our SentiWeibo dataset construction begins with hashtags, which serve as concise and effective identifiers encapsulating the core themes and key information of specific topics of social discussions~\citep{Laucuka2018functionofhashtags}. These hashtags are sourced from two primary channels, ensuring diversity and comprehensive coverage of trends. 
The first channel is Weibo Hot Topics, a reliable indicator of trending themes within news content. From this source, we selected a substantial portion of the dataset, encompassing a wide range of topics such as entertainment, sports, societal issues, and daily life. 
The other channel is the Weibo AI Search platform, an AI-powered tool that highlights both trending and emerging hashtags. Through these dual channels, we curated a diverse set of 53 hashtags that capture both dominant social discussions and emerging trends.

For each hashtag, we collect hundreds of posts with hashtag along with their corresponding public sentiment distributions provided by Weibo, which will then be used to calculate ground truth sentiment distribution.
Additionally, for each user who contributed to these discussions, we retrieve their historical posts to construct comprehensive user profiles, enabling analysis of their behavioral patterns and viewpoint evolution.

\subsection{Data Pre-processing}

The directly collected dataset inherently includes a considerable amount of noise, which could hinder meaningful analysis. 
To address this issue, we employed a systematic data-cleaning protocol involving human annotators who manually evaluated each hashtagged post for relevance and quality. 
Posts that lacked thematic alignment with their associated hashtags were excluded to ensure the dataset remained focused and cohesive. 
Furthermore, content originating from mainstream media outlets was omitted, as such posts often represent neutral reportage rather than the personal perspectives critical to our research objectives.
This meticulous manual pre-processing ensures that the dataset maintains its authenticity and aligns with the goals of our study.

After initial data cleaning, we analyzed users' historical posts, which still contained substantial noise—such as giveaway reposts or daily life updates—that offered little analytical value for studying public discourse. Given the large volume of data, a full manual review was infeasible due to time and resource constraints.
Empirical analysis revealed that posts containing hashtags tend to convey higher information density and clearer event-specific opinions. To address this, we developed a two-step filtering strategy: (1) retaining posts with hashtags as a signal of topical relevance, and (2) removing posts containing noise-indicating keywords.
These keywords (e.g., “fashionable outfit”, “positive recommendation”) were derived through a systematic content analysis of 1,000 posts, identifying patterns strongly associated with low informational value even in hashtagged content. 
This dual-filtering approach effectively removed noise while preserving meaningful user perspectives.
To validate the method, we conducted a human evaluation on 100 randomly sampled cases, with only 1\% flagged as irrelevant or noisy—demonstrating high filtering precision. 

The final dataset consists of 53 topics with official sentiment distributions,  7,837 hashtagged posts from 6,401 unique users, along with 476,605 historical posts, where each user contributes between 10 and 953 entries, forming a rich, longitudinal corpus for analysis.
We split the dataset into training, validation, and test sets using an 8/1/1 ratio.

\section{Method}

% In this section, we introduce our proposed SocialAlign framework in detail for the public response prediction task.

\begin{figure*}[tb]
    \centering
    \includegraphics[width=0.9\linewidth]{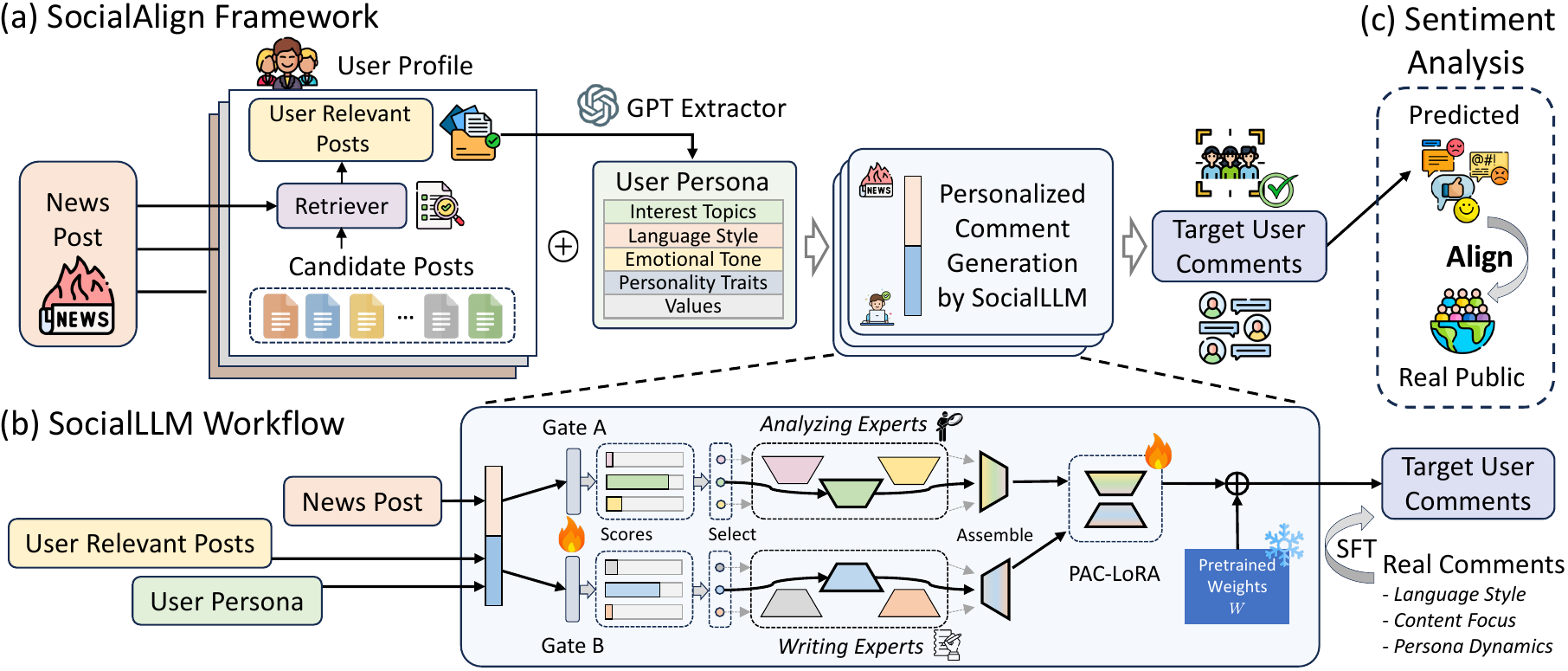}
    \caption{
SocialAlign Framework.
(a) The pipeline retrieves users' historical posts and extracts personas from predefined aspects.
(b) During SocialLLM training, news and user data dynamically select analyzing and writing experts for personalized comment generation.
(c) Dual-level analysis compares predicted and real comments for sentiment alignment and group distribution.
    }
    \label{fig:model}
\end{figure*}

% \subsection{Framework Overview}
SocialAlign unifies personalized response generation and group-level sentiment analysis.
As shown in Figure~\ref{fig:model}, it encodes news, retrieves user history, and builds five-dimensional personas.
SocialLLM activates expert modules to generate responses, which are aggregated into public sentiment distributions.
% A dual-level analysis ensures alignment with both individual perspectives and real-world sentiment trends.

% This section details SocialAlign’s three stages: input processing, personalized generation, and macro-level sentiment aggregation.

\subsection{Persona Construction for SocialLLM}
User personas have proven to be powerful for enhancing personalization across various tasks. For instance,~\citet{sugiyama2004adaptive} demonstrated that automatically constructed user personas can achieve highly personalized search results without requiring explicit user input, while~\citet{ouaftouh2019social} established the effectiveness of persona-based clustering in recommendation systems. Motivated by their success, we propose a novel persona-driven approach for personalized comment generation.

% 方法的分析视角+例子
As shown in Figure~\ref{fig:model}(a), our approach begins by constructing rich user personas to guide SocialLLM in generating personalized, context-aware responses. Unlike traditional methods that rely solely on historical posts, our method captures deeper, latent patterns through systematic analysis of 1,000 users’ reactions to social events. We identify five key persona dimensions: interests (e.g., housing, work life, social justice), language style (e.g., sarcasm, humor, directness), emotional tone (e.g., negativity, anger), personality traits (e.g., critical thinking, pragmatism), and values (e.g., social equity, concern for the marginalized).
For instance, in response to the news topic \#New Mortgage Policy Promises Minor Savings on Long-term Loans\#, a user comments: \textit{“Why must we take bank loans and pay interest for housing? Whose pockets are we filling?”} This reveals the user’s \textit{interest} in housing issues, a direct and questioning \textit{language style}, a negative \textit{emotional tone} toward financial institutions, a \textit{critical-thinking personality}, and \textit{values} emphasizing market fairness.
When presented with a related news topic \#Shanghai Housing Prices Surge Dramatically Overnight\#, the same user responds: \textit{“This is insane! Houses are for living, not speculation—return to normal commodity value!”} This comment reflects the same consistent \textit{persona dimensions}, confirming the model’s ability to capture and generalize user-specific attitudes across different yet related contexts.

% 具体实现
Specifically, our persona construction process consists of three steps. First, for each news post \( x^\mathcal{N} \) and user \( u \), we employ BM25~\citep{robertson2009probabilistic} to retrieve the top-\( k \) relevant documents from the user's historical posts. Next, we utilize GPT-4~\citep{openai2024gpt4technicalreport} to analyze these documents and generate structured personas $p_u$ along the five identified dimensions. Finally, the retrieved history $\mathcal{H}_u^R$ and constructed persona $p_u$ are fed into SocialLLM to generate the personalized response:

\begin{equation}
   y_u^{\prime \mathcal{N}} = \text{SocialLLM}(\mathcal{H}_u^R, p_u, x^\mathcal{N}).
\end{equation}

\subsection{SocialLLM Architecture Design}
In this subsection, we introduce SocialLLM, a specialized LLM fine-tuned using the novel Personalized Analyze-Compose Low-Rank Adaptation (PAC-LoRA) architecture, as illustrated in Figure~\ref{fig:model}(b).
% SocialLLM serves as the core component of the SocialAlign framework and dynamically adjusts its response generation to news events by leveraging user-specific information.
% This approach enables the model to generate outputs precisely tailored to individual user profiles and their historical interactions.

\subsubsection{Personalization Strategy}
Motivated by real-world observations, we conceptualize the user response generation as a two-stage cognitive process: i) decomposition and analysis of key information from news content, and ii) personalized reasoning to formulate individual perspectives.  This process recognizes that different news topics demand distinct analytical approaches, while individual users exhibit varying response patterns and attitudes toward the same content. Guided by this understanding, we design our model with specialized expert modules that dynamically adapt to diverse content types and user characteristics, enabling context-aware and personalized response generation.

\subsubsection{Preliminary LoRA Structure}
Building on the above strategy, we begin by introducing the preliminary LoRA~\citep{hu2022lora} structure before detailing its adaptation into PAC-LoRA. In LoRA, a pre-trained weight matrix $\mathbf{W}_0 \in \mathbb{R}^{d \times k}$ is updated using a low-rank decomposition consisting of $\mathbf{A}$ and $\mathbf{B}$, where $\mathbf{A} \in \mathbb{R}^{r \times k}$ and  $\mathbf{B} \in \mathbb{R}^{d \times r}$, with $r \ll \min(d, k)$.
During the training, \( \mathbf{W}_0 \) is frozen and does not receive gradient updates, while \( \mathbf{A} \) and \( \mathbf{B} \) contain the trainable parameters.
Given \( \mathbf{h} = \mathbf{W}_0 \mathbf{x} \), the modified forward pass becomes:
\begin{align}
    \mathbf{h}' = \mathbf{W}_0 \mathbf{x} + \Delta \mathbf{W} \mathbf{x} = \mathbf{W}_0 \mathbf{x} + \mathbf{B} \mathbf{A} \mathbf{x}.
\end{align}

In this framework, $\mathbf{A}$ is responsible for learning compact representations of the input, while $\mathbf{B}$ decodes these representations into coherent outputs.
To align with the response-writing process, we reinterpret $\mathbf{A}$ as an \textit{analyzing matrix} and $\mathbf{B}$ as a \textit{writing matrix}, providing a more intuitive understanding of the model structure.

We design SocialLLM to generate personalized responses, adapting to diverse user-specific reasoning and understanding styles. To this end, we extend LoRA into PAC-LoRA by introducing multiple analyzing and writing experts. Each user is assigned an adaptive distribution over these experts, enabling the model to generate outputs dynamically tailored to individual user profiles.

\subsubsection{PAC-LoRA Structure}
Concretely, PAC-LoRA consists of multiple analyzing experts $\mathbf{A}_i$ and writing experts $\mathbf{B}_i$.
During training, the pre-trained weights  $\mathbf{W}_0$  remain frozen, while the experts  $\mathbf{A}_i$  and  $\mathbf{B}_i$  are optimized. 
The forward propagation in PAC-LoRA is modified to include a low-rank adjustment:
\begin{equation}
    \mathbf{h}^{\prime} = \mathbf{W}_0 \mathbf{x} + \Delta \mathbf{W} \mathbf{x} = \mathbf{W}_0 \mathbf{x} + \textstyle \sum_{i=1}^ {N_a} (\mathbf{g}_i^B \mathbf{B}_i)(\mathbf{g}_i^A \mathbf{A}_i \mathbf{x}),
\end{equation}
where $N_a$ represent the expert number, \( \mathbf{g}_i^A \) and \( \mathbf{g}_i^B \) are the gating weights applied to the \( i \)-th analyzing expert \( \mathbf{A}_i \) and writing expert \( \mathbf{B}_i \), respectively. 
These gating weights are computed via a gating function \( \mathbf{G} \), implemented using a multi-layer perceptron:
\begin{equation}
\mathbf{g}_i^A = \text{softmax}(\mathbf{G}^A(\mathbf{x}_{\text{news}}))_i, \quad \mathbf{g}_i^B = \text{softmax}(\mathbf{G}^B(\mathbf{x}_{\text{user}}))_i,
\end{equation}
where $\mathbf{x}_{\text{news}}$ and $\mathbf{x}_{\text{user}}$ denote inputs for news content and user-specific features.
The dynamically computed gating weights  $\mathbf{g}_i^A$  and  $\mathbf{g}_i^B$ guide the selection of experts  $\mathbf{A}_i$ and  $\mathbf{B}_i$. 
This mechanism allows \( \mathbf{A}_i \) to specialize in capturing diverse topics from the news content, while \( \mathbf{B}_i \) encodes personalized reasoning and response styles. 
% By dynamically assigning weights based on news content and user-specific features, PAC-LoRA achieves a balance between generalization across topics and personalization for individual users.

\subsection{Macro-level Sentiment Prediction}

Our task involves predicting responses at both the micro and macro levels. At the micro level, the focus is on personalized comment generation, while at the macro level, the goal is to predict group sentiment distributions to capture collective public attitudes.

Previous works often rely on fine-tuned smaller models like BERT~\citep{devlin2019bert}, trained on general-purpose text corpora such as news or Wikipedia. 
However, social media language is highly dynamic and evolving, incorporating elements such as emoticons, internet slang, and implicit expressions, all of which require a deeper understanding of the social context.
As a result, conventional sentiment classification models struggle to adapt to these unique linguistic characteristics and the rapid evolution of social media language~\citep{SANCHEZRADA2019socialcontext}.

For domain adaptation, we adopt Qwen-Max, which pre-trained on massive internet corpora, to classify each user's generated comment \(r^{\prime \mathcal{N}}_u\) and real comment \(r^{\mathcal{N}}_u\) into one of seven sentiment categories derived from the emotion taxonomy provided by Weibo:
\begin{align}
    s^{\prime}_u = \text{Qwen-Max}(y_u^{\prime \mathcal{N}}) \in \mathcal{S},
\end{align}
where $\mathcal{S}$ denotes the set of emotion labels \{\textit{happy, sad, angry, calm, fear, surprised, disgusted}\}.

Finally, at the macro level, we aggregate individual sentiments \(\{s^{\prime}_1, \ldots, s^{\prime}_M\}\) to form \(P^{\prime}\), the predicted sentiment distribution, and compare it to the ground truth distribution \(P\) for each topic.

\section{Experiments}
\subsection{Implementation Details}  
We developed a web spider utilizing the Scrapy and Selenium frameworks to collect Weibo posts.  
For the model architecture, we selected Qwen2.5-7B-Instruct~\citep{qwen2}, a specialized instruction-following model, as the base model for SocialLLM.
All models are fine-tuned on a single NVIDIA RTX 6000 Ada GPU.
% To optimize computational costs, we employed BF16 precision along with a gradient accumulation strategy.
We set the number of experts, \( N_a \), to three. 
All models were trained with a learning rate of \(2 \times 10^{-5}\). PCGN adopted a batch size of 4, while other models used a batch size of 1. For LoRA-based models, the rank was set to 8, and gradient accumulation was set to 16 (8 for PCGN). 
The training process spanned 3 epochs for PCGN and 2 epochs for other models. 
We utilized the AdamW optimizer~\citep{loshchilov2019decoupledweightdecay} with a dropout rate of 0.1. 
For user persona extraction and sentiment classification, we utilized GPT-4-0125-preview and Qwen-Max, respectively.

\subsection{Baselines}
We evaluate SocialAlign against a range of baselines, including PLM, LLM, and fine-tuned LLMs.
First, we adopt a naive baseline:
(1) QWen-SP (sentiment predictor), which reads the target news as well as retrieved posts, and directly predicts sentiment.
We also include several few-shot LLM baselines that are provided with retrieved historical user posts:
(2) LLaMA2-7B-Chat \citep{touvron2023llama2},
(3) LLaMA3-8B-Chat \citep{dubey2024llama}, and
(4) Qwen2.5-7B-Instruct \citep{yang2024qwen2}.
For fine-tuned baselines, we consider:
(5) PCGN \citep{zeng2019automatic}, a model originally designed for comment generation.
To enhance its performance, we replace its basic LSTM \citep{lstm} backbone with BART-large-Chinese \citep{lewis2020bart, Shao2021cpt} and fine-tune it on the SentiWeibo dataset.
Next, we explore different fine-tuning strategies using the same backbone—Qwen2.5-7B-Instruct—which we found to outperform the LLaMA-based models:
(6) LoRA \citep{hu2022lora}, a parameter-efficient fine-tuning method, is used to adapt Qwen2.5 to SentiWeibo.
(7) PER-PCS \citep{tan2024personalized}, a personalization-oriented method, trains separate LoRA modules for different users and combines them to generate responses for unseen users.
However, given the scale of our user base, training a unique LoRA module for each user is impractical. To address this, we cluster users into a single group and train three LoRA modules accordingly.
Our proposed PAC-LoRA is also built on the Qwen2.5-7B-Instruct backbone.
To ensure a fair comparison, all models are provided with the same retrieved historical posts as input.

\begin{table*}[tb]
\centering
\small
\begin{tabular}{c|l|c|c|c|c|c|c|c}
\toprule
\textbf{Topic} & \textbf{Metric} & \textbf{Qwen-SP} & \textbf{Llama} & \textbf{Qwen} & \textbf{PCGN} & \textbf{LoRA} & \textbf{PER-PCS} & \textbf{PAC-LoRA (Ours)} \\
\hline
\multirow{7}{*}{Public Health} & \multicolumn{6}{@{}l}{\emph{  Micro-level:}} \\
 & Sentiment Acc $\uparrow$ & 8.4 & 17.1 & 11.7 & 12.7 & 27.8 & 27.8 & \textbf{35.6} \\
 & Sentiment F1 $\uparrow$ & 9.7 & 15.5 & 9.1 & 13.5 & 34.3 & 32.6 & \textbf{38.5} \\
 & Comment $S_{\text{LLM}}$ $\uparrow$ & - & 13.9 & 36.1 & 24.0 & 37.1 & 37.3 & \textbf{40.8} \\
 & Comment $S_{\text{Human}}$ $\uparrow$ & - & 9.9 & 28.3 & 16.0 & 32.4 & 33.6 & \textbf{41.6} \\
 & \multicolumn{6}{@{}l}{\emph{ Macro-level:}} \\
 & Group-level JS Div $\downarrow$ & 65.9 & 50.3 & 56.2 & 58.6 & 35.4 & 39.3 & \textbf{34.1} \\
\hline
\multirow{7}{*}{Recruitment Policies} & \multicolumn{6}{@{}l}{\emph{ Micro-level:}} \\
 & Sentiment Acc $\uparrow$ & 18.8 & 23.9 & 8.0 & 20.3 & 27.5 & 29.0 & \textbf{43.5} \\
 & Sentiment F1 $\uparrow$ & 12.3 & 30.6 & 9.9 & 25.6 & 36.8 & 34.3 & \textbf{50.2} \\
 & Comment $S_{\text{LLM}}$ $\uparrow$ & - & 7.3 & 27.4 & 16.4 & 33.7 & 31.7 & \textbf{38.3} \\
 & Comment $S_{\text{Human}}$ $\uparrow$ & - & 7.9 & 13.5 & 12.5 & 29.1 & 30.9 & \textbf{39.2} \\
 & \multicolumn{6}{@{}l}{\emph{Macro-level:}} \\
 & Group-level JS Div $\downarrow$ & 60.6 & 46.5 & 67.5 & 51.9 & 40.5 & 40.7 & \textbf{26.1} \\
\hline
\multirow{7}{*}{Financial Scams} & \multicolumn{6}{@{}l}{\emph{ Micro-level:}} \\
 & Sentiment Acc $\uparrow$ & 32.9 & 24.0 & 37.7 & 37.3 & 44.4 & 42.4 & \textbf{48.2} \\
 & Sentiment F1 $\uparrow$ & 34.0 & 30.9 & 42.6 & 43.3 & 47.5 & 46.9 & \textbf{50.1} \\
 & Comment $S_{\text{LLM}}$ $\uparrow$ & - & 13.3 & 37.9 & 28.0 & 46.8 & 48.1 & \textbf{48.9} \\
 & Comment $S_{\text{Human}}$ $\uparrow$ & - & 7.9 & 22.4 & 13.6 & 35.9 & 32.7 & \textbf{38.8} \\
 & \multicolumn{6}{@{}l}{\emph{ Macro-level:}} \\
 & Group-level JS Div $\downarrow$ & 42.2 & 40.4 & 22.2 & 38.9 & 22.8 & 23.5 & \textbf{17.5} \\
\bottomrule
\end{tabular}
\caption{The performance of our PAC-LoRA model and baselines on public sentiment prediction is evaluated across three different topics. 
% The evaluation includes micro-level sentiment label accuracy, quality assessment of generated comments, and macro-level sentiment label distribution accuracy.
Numbers in \textbf{bold} mean that the improvement to the best baseline is statistically significant (a two-tailed paired t-test with p-value $<$ 0.01).
All results are normalized to a 0-100 scale.}
\vspace{-3mm}
\label{tab:main}
\end{table*}

\begin{table}[tbp]
\centering
\small
\begin{tabular}{@{}c|cccccc@{}}
\toprule
\multirow{2}{*}{Model} & \multicolumn{2}{c}{Language Style} & \multicolumn{2}{c}{Content Focus} & \multicolumn{2}{c}{Persona Dynamics} \\ 
 \cmidrule(r){2-3} \cmidrule(r){4-5} \cmidrule(r){6-7}
 & $S_{\text{LLM}}$ & $S_{\text{Human}}$ & $S_{\text{LLM}}$ & $S_{\text{Human}}$  & $S_{\text{LLM}}$ & $S_{\text{Human}}$ \\
\midrule
 Llama & 9.1 & 6.7 & 17.0 & 10.3 & 11.0 & 8.7\\
% \hline
Qwen & 31.5 & 21.7 & 45.7 & 23.3 & 29.3 & 19.1\\
PCGN & 25.5 & 15.5 & 29.4 & 13.6 & 19.5 & 13.1\\
LoRA & 37.9 & 33.3 & 49.2 & 32.3 & 35.1 & 31.7\\
PER-PCS & 39.1 & 33.2 & 48.4 & 32.1 & 38.9 & 32.0\\
PAC-LoRA & \textbf{40.9}  & \textbf{40.1} & \textbf{50.7} & \textbf{39.7} & \textbf{42.7} & \textbf{39.7}\\
\bottomrule
\end{tabular}
\caption{Dimension scores for comment generation.
\textbf{Bold} numbers indicate statistically significant improvements over the best baseline (two-tailed paired t-test, $p$\textless0.01). }
\label{tab:human}
\vspace{-8mm}
\end{table}

\subsection{Metrics}
% We evaluate our framework at both micro and macro levels.

At the micro level, we evaluate individual responses from two aspects: sentiment alignment and personalized comment generation.
For \textbf{sentiment alignment}, we utilize authentic user comments as ground-truth labels, computing topic-wise overall accuracy and F1 score to evaluate the consistency between model predictions and actual sentiments. 
For \textbf{comment generation}, we adopt a relative scoring scheme on a 0–10 scale that reflects the degree to which the generated comment aligns with the user's profile.
The scoring considers three key dimensions derived from our profiling method:
\textit{i) Language Style} corresponds directly to the same aspect in user profiling, evaluating whether the generated comments align with a user’s characteristic expression patterns, such as sarcasm, humor, or directness, as a primary indicator of personalization quality. 
\textit{ii) Content Focus}, synthesizing Interests and Values, assesses whether the generated comments address topics the user consistently engages with (e.g., housing market issues) and reflect their value orientations (e.g., advocacy for fair market practices). 
\textit{iii) Persona Dynamics}, combining Personality Traits and Emotional Tone, evaluates whether the generated comments capture the user’s characteristic thinking patterns (e.g., critical thinking) and emotional tendencies (e.g., skepticism toward financial institutions).
To implement this metric, we employ both automated and human evaluations. 
For automated assessments, we leverage Qwen-Max as a judge, taking advantage of its strong alignment with human preferences~\cite{zheng2024llmasjudge}.
To mitigate potential deviations in LLM-based evaluations, we supplement this with a human evaluation with three annotators.
Each annotator is presented with original user comments with the corresponding generated responses from different models and rates the responses on a 0-10 scale using the same criteria. 

At the macro level, we use Jensen-Shannon (JS) divergence to measure alignment between predicted and real public sentiment distributions, where lower values indicate better reflection of real-world attitudes.

\subsection{Main Result}
To comprehensively evaluate SocialAlign's alignment capability with real-world responses on social events, we randomly select three trending hashtags from Weibo during November 2024: Public Health (\#Tourists Poisoned Due to the Guide Using Ethylene Glycol as Hotpot Soup\#), Recruitment Policies (\#Ban on Limiting Campus Recruitment to Top Universities\#), and Financial Scams (\#Revelation of a High-Paying Job Concealing a Medical Beauty Scam\#).
These hashtags encompass a diverse range of social topics, enabling a robust analysis of sentiment alignment.
The dataset consists of the official sentiment distribution of each topic, 731 users, 751 hashtag posts, with 47,502 history posts, providing a comprehensive view of how these events are perceived and discussed online.
The results are shown in Table~\ref{tab:main} and Table~\ref{tab:human}.

\begin{figure*}[htb]
    \centering
    \includegraphics[width=0.85\linewidth]{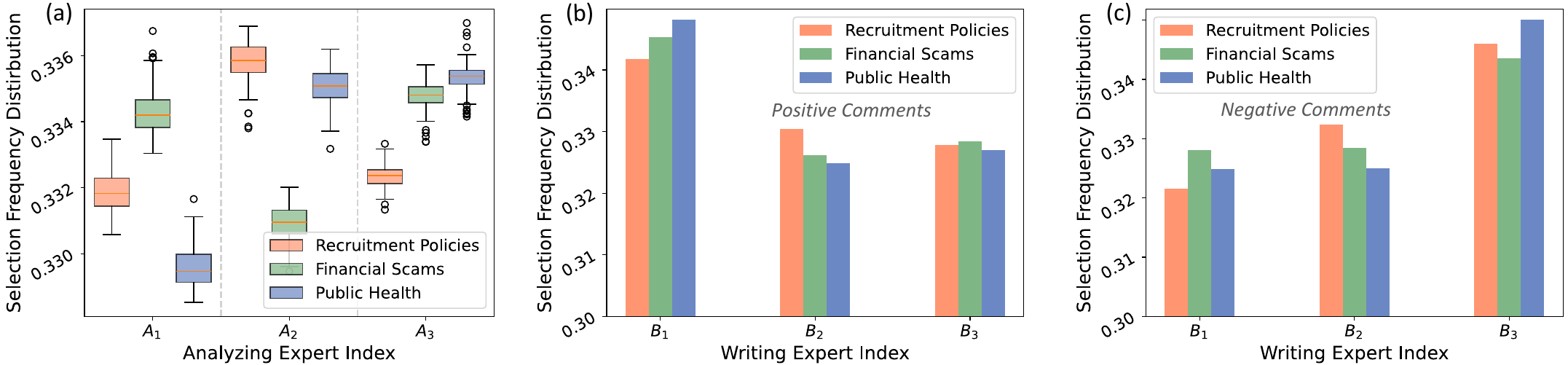}

    \caption{
   Analysis of our PAC-LoRA structure.
(a) Preferences of three analyzing experts for different news content across all comments.
(b, c) Utilization ratios of three writing experts for three randomly sampled (b) Positive and (c) Negative comments.
    }
    \label{fig:analysis}
\end{figure*}

Our first observation is that \textit{personalization is fundamentally different from general-domain tasks and requires specialized fine-tuning}.
For example, Qwen, despite its larger size, consistently underperforms compared to the smaller PCGN model, especially on the Recruitment Policies topic. Llama2 and Llama3 show similar issues, often producing mixed-language outputs with nonsensical fragments, leading to very low $S_{\text{LLM}}$ and $S_{\text{Human}}$ scores. For simplicity, we report Llama2 as a representative case.
These results show that model size alone is not enough—effective personalization depends on domain-specific fine-tuning. In contrast, fine-tuned models like LoRA and PER-PCS perform significantly better, reinforcing the value of targeted adaptation.

Secondly, \textit{micro-level and macro-level evaluations assess the task from different perspectives}. 
For instance, PER-PCS achieves better micro-level performance, demonstrating strong personalized response generation quality and sentiment alignment. 
However, at the macro level, it struggles to align with public sentiment distributions, as reflected in its higher JS divergence.
Similarly, as described earlier, Llama2 faces a comparable issue: despite achieving high sentiment accuracy, its comment quality remains poor. 
This discrepancy underscores the importance of considering both levels of evaluation for a comprehensive assessment.

Thirdly, \textit{existing evaluation LLMs demonstrate relatively good performance but also have limitations when evaluating posts on social platforms}. 
Comparing $S_{\text{LLM}}$ and $S_{\text{Human}}$, we observe that they align in most cases, where models with higher $S_{\text{LLM}}$ also tend to have higher $S_{\text{Human}}$. 
The correlation between \( S_{\text{LLM}} \) and \( S_{\text{Human}} \) reaches as high as 0.89 across all topics (p < 0.01), demonstrating a strong and statistically significant alignment between model-generated and human-written comments.
However, human evaluation reveals a more significant performance gap between models, while the differentiation ability of LLMs is limited, resulting in a narrower performance gap.
Note that the Kappa scores for the three dimensions are 0.47, 0.44, and 0.49, respectively, indicating a moderate level of agreement among the three annotators~\cite{chmura2002kappa}.
This demonstrates that the annotators share a reasonable level of consensus, reflecting a shared understanding of the task.

Finally, \textit{Our model, PAC-LoRA, achieves the best performance across all three topics and excels in both micro-level and macro-level evaluations, particularly in human evaluation}. This observation demonstrates that our proposed framework excels in both personalized comment generation quality and public sentiment alignment.

\section{Analysis and Discussion}

\subsection{Ablation Study}

We conduct an ablation study to examine the contribution of each key component in our framework, as shown in Table~\ref{tab:ablation}. 
For model architecture, removing the gating mechanism of multiple analyzing experts (w/o Analyzing Experts) leads to the accuracy falling to 30.4\%. Similarly, replacing the selection of writing experts with uniform weight combinations (w/o Writing Experts) makes the accuracy decrease from 43.9 to 30.4. 
% Specifically, sentiment accuracy drops from 43.9\% to 32.8\% and 30.4\%, respectively. 
This highlights the crucial role of the adaptive selection process for multi-analyzing and writing experts.
When user historical posts and personas were removed, the performance also declined. 
Notably, eliminating history posts (w/o Historical Post) causes a larger drop compared to removing personas (w/o Persona), which shows user history serves as a prime reference for predicting user comments, while our proposed persona effectively supplements users' personal information.
Overall, these findings validate the necessity of each component in achieving optimal performance.

\begin{table}[tbp]
\centering
\small
\begin{tabular}{@{}l|ccccc@{}}
\toprule
& Acc & F1 & $S_\text{LLM}$ & $S_\text{Human}$ & JS Div   \\ \midrule
Full Architecture  & \textbf{43.9} &  \textbf{48.2} & \textbf{44.7} & \textbf{39.9} & \textbf{20.2} \\
% \hline
-w/o Analyzing Experts  & 32.8 & 39.8 & 43.3 & 38.6 & 30.1 \\ 
-w/o Writing Experts & 30.4 & 37.5 & 42.2 & 37.5 & 34.0 \\ 
-w/o Historical Post & 30.0 & 37.4 & 41.4 & 32.0 & 33.9 \\ 
-w/o Persona & 33.3 & 40.2 & 41.6  & 32.3 & 30.1 \\ 
\bottomrule
\end{tabular}
\caption{Ablation study on our SocialAlign framework.
\textbf{Bold} numbers indicate statistically significant improvements over the best baseline (two-tailed paired t-test, $p$\textless0.01). }
\label{tab:ablation}
\end{table}

\subsection{Generalization Evaluation}
In addition to evaluating on our own dataset, we also assess the generalizability of our PAC-LoRA by adapting it to the LaMP~\cite{chen2024large} benchmark. Specifically, we focus on the personalized news headline generation task (LaMP-4), where the goal is to generate a headline for a news article based on its content and the historical writing style of a specific user.
For a fair comparison, we employ Qwen2.5-7B-Instruct as backbone and utilize BM25 to retrieve one relevant historical sample from the user profile, which is used as context input for all models. 
As shown in Table~\ref{tab:lamp}, our PAC-LoRA component consistently outperforms other adapted vanilla model and LoRA variants across the evaluated metrics, demonstrating strong generalization across datasets and tasks.

\begin{table}[tbp]
\centering
\small
\begin{tabular}{@{}c|cccc@{}}
\toprule
Metric & Vanilla & LoRA & PER-PCS & PAC-LoRA   \\ \midrule
ROUGE-1  & 15.17 &  19.13 & 19.30 & \textbf{19.68}  \\
ROUGE-2  & 3.87 & 6.46 & 6.59 & \textbf{6.65}  \\ 
ROUGE-L  & 13.36 & 17.27 & 17.53 & \textbf{17.70}  \\ 
\bottomrule
\end{tabular}
\caption{Personalized news headline generation on LaMP benchmark.
\textbf{Bold} numbers indicate statistically significant improvements over the best baseline. 
All results are normalized to a 0-100 scale.}
% \vspace{-10mm}
\label{tab:lamp}
\end{table}

\subsection{Expert Utilization Analysis of PAC-LoRA}

PAC-LoRA integrates two types of experts: analyzing experts for understanding news content and writing experts for generating personalized responses. This section examines how effectively the model leverages these experts for different types of content.

\textbf{Utilization of Analyzing Experts.}  
We randomly select one intermediate layer from SocialLLM and analyze the utilization ratios of analyzing experts across all users for each topic. 
As shown in Figure~\ref{fig:analysis}(a), distinct usage patterns emerge.  
For instance, the topics \textit{Public Health} and \textit{Recruitment Policies} exhibit a higher utilization of expert \( \mathbf{A}_2 \). 
This observation suggests that \( \mathbf{A}_2 \) is effective at capturing features related to collective societal concerns, such as health risks and fairness in employment opportunities, both of which require an understanding of broad public discourse and policy implications.  
In contrast, \textit{Financial Scams} and \textit{Public Health} show higher usage of expert \( \mathbf{A}_3 \), suggesting that it encodes features related to individual harm and unethical practices, such as scams or health threats.

% For \textit{Public Health} and \textit{Recruitment Policies}, expert \( \mathbf{A}_2 \) is most frequently used, indicating its strength in capturing collective social concerns such as health risks and employment fairness.  
% In contrast, \textit{Financial Scams} and \textit{Public Health} show higher usage of expert \( \mathbf{A}_3 \), suggesting that it encodes features related to individual harm and unethical practices, such as scams or health threats.

\textbf{Utilization of Writing Experts.}
Writing experts operate at the individual comment level. Figures~\ref{fig:analysis}(b) and (c) show their usage across three positive and three negative comments, respectively.
In (b), $\mathbf{B}_1$ dominates in positive comments. For example, a user praises recruitment policies: "Feels pretty good, it shows that our talent perspective has improved... A great trend of change." Another warns against scams: "Scams are so common—everyone must be cautious!" In public health: "Luckily they were saved... Don’t mess with others’ stuff." These optimistic tones align with $\mathbf{B}_1$.
In (c), $\mathbf{B}_3$ appears most in negative comments. One frustrated job-seeker writes: "I was shocked... Am I unsuitable just because my alma mater isn't prestigious?" On scams: "Still taking loans despite knowing it’s a scam... that’s gambling with your future." On public health: "Do things like this really happen? What is going on..." These critical or skeptical tones reflect $\mathbf{B}_3$’s role.
These results show that writing experts dynamically adapt to user sentiment, enabling nuanced, context-aware response generation.
% B\_1 comments:

% Recruitment Policies: Feels pretty good, it shows that our talent perspective has improved a lot. We no longer confine our focus to a single school or region but instead value a person's abilities and potential more. This is a great trend of change, and I hope that more companies and institutions will be able to select talent more fairly and impartially in the future."

% Financial Scams: Scams are so common, so everyone must be cautious to prevent being scammed!!!

% Public Health: Luckily, they were saved... In the future, be careful and don't tamper with others' things casually~

% B\_3 comments:

% Recruitment Policies: I once interviewed for a position at a state-owned enterprise, and the HR directly told me, "We don't particularly like students from your university." I was shocked and thought to myself: Does being such an excellent student mean I'm unsuitable for your company just because my alma mater isn't prestigious enough? Later, I realized that some state-owned enterprises place a high value on prestigious schools. If they don't urgently need talent, they would rather leave a position vacant than accept students from second or third-tier universities. This mentality is truly frightening!

% Financial Scams: This is truly a human tragedy! Knowing you're being scammed yet continuing to take out loans for cosmetic procedures is just taking small risks for huge losses.

% Public Health: Do such accidents really happen in real life? What on earth is this...

\subsection{Case Study}

\begin{figure}[tbp]
    \centering
    \includegraphics[width=0.9\linewidth]{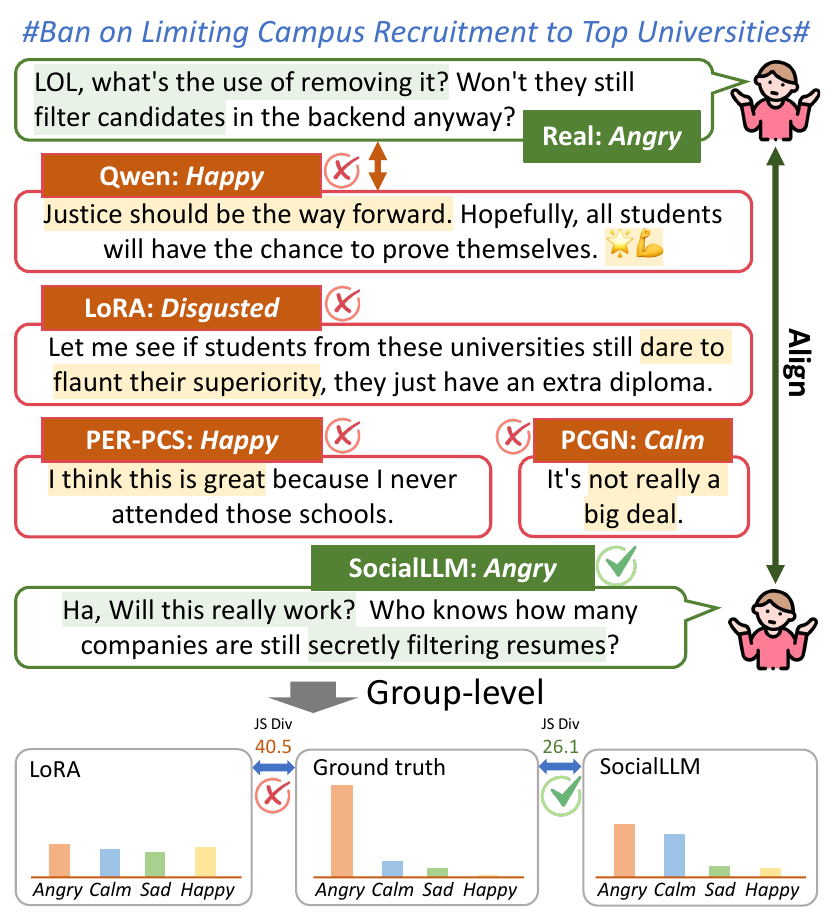}
    \caption{Comparison of responses generated by SocialAlign and different baselines.}
    \vspace{-3mm}
    \label{fig:response_case}
\end{figure}
We also give a case study in Figure~\ref{fig:response_case} to provide a more intuitive understanding.
It can be observed that \textit{LLMs not specifically trained for personalization tasks, such as vanilla Qwen, tend to express overly optimistic and politically correct opinions, lacking both social context awareness and user-specific language style}.
In contrast, more adapted baselines fine-tuned on our SentiWeibo dataset are able to discuss events in a manner resembling a real person. For example, comments generated by PER-PCS include personal experiences (e.g., "I never attended those schools"). However, it still misses certain nuanced viewpoints, such as the skepticism towards policy implementation that characterizes authentic discussions. \textit{Only our SocialAlign effectively captures authentic users' outspoken linguistic styles and sarcastic emotional tendencies}, as demonstrated by phrases like "will this really work?", while simultaneously maintaining the user's disdain for falsehoods and insincerity, as seen in comments like "secretly filtering resumes." 
Built on the well-generated comment, the group-level sentiment distribution of our model is also more accurate compared to the baselines.
This highlights the superior ability of SocialAlign to align with real-world social contexts and individual user preferences.

\section{Conclusion}

We present SocialAlign, the first framework to align large language models with public sentiment at both micro and macro levels via personalized comment generation and group sentiment modeling.
Our model, SocialLLM, captures individual user traits while staying aligned with broader societal sentiment.
We curate a rich dataset from social platforms, including user histories, profiles, and sentiment distributions—often missing in prior work.
Experiments show that SocialAlign enhances response generation by balancing personalization and public alignment.
Future work includes extending to multilingual settings, adding multimodal inputs, and modeling evolving user sentiment over time.

% \subsection*{Ethics Discussion}

% This section outlines the ethical considerations of using Weibo data.
% Firstly, Weibo is widely used in academic research, with many studies releasing datasets for tasks like question answering, opinion analysis, fake news detection, and propaganda research~\citep{fu2013assessing,fu2023propagandization,gan2022wuhandiary,hu2024bad}.
% Similarly, platforms like Twitter have been extensively used in analogous studies~\citep{wagh2018survey,alqurashi2020large}, establishing a strong precedent for utilizing public social media data in research.
% Secondly, Weibo provides API access for developers to collect publicly available data in compliance with its terms of service. 
% Researchers can retrieve public posts, user profiles, and metadata using authorized API keys ensuring that data collection follows platform policies.
% To address further ethical concerns, personally identifiable information was anonymized and only publicly available data was utilized.
% The data collection process complies with relevant legal and ethical standards, including data protection regulations like GDPR where applicable.
% To support transparency and reproducibility, we release both the data collection and model code.

\subsection*{Ethics Discussion}

We carefully considered ethical aspects of using Weibo data. Weibo is commonly used in academic research with numerous studies releasing datasets for various tasks~\citep{fu2013assessing,fu2023propagandization,gan2022wuhandiary,hu2024bad}, similar to Twitter usage in comparable studies~\citep{wagh2018survey,alqurashi2020large}. Our data collection follows Weibo's terms of service through authorized API access. All personally identifiable information was anonymized, with only publicly available data utilized in compliance with relevant data protection regulations. For transparency and reproducibility, we release both our data collection methodology and model code.
% In summary, this study upholds responsible and ethical use of Weibo data by following established research practices, respecting platform policies, and implementing robust privacy safeguards.

\bibliographystyle{ACM-Reference-Format}
\bibliography{main}

\end{document}